\documentclass{iaus}
\usepackage{graphicx}
\usepackage[authoryear, round]{natbib}
\usepackage{url}

\title[Guide to Simulation:Observation Comparison] %% give here short title %%
{A Guide to Comparisons of Star Formation Simulations with Observations}

\author[Alyssa A. Goodman]   %% give here short author list %%
{Alyssa A. Goodman$^1$}

\affiliation{$^1$Harvard-Smithsonian Center for Astrophysics \\ 60 Garden Street, MS 42\\ Cambridge, MA 02318, USA\\ email: {\tt agoodman@cfa.harvard.edu} }
\pubyear{2010}
\volume{170}  %% insert here IAU Symposium No.
\pagerange{xxx}
\jname{Computational Star Formation}
\editors{J. Alves, B. Elmegreen \& V. Trimble, eds.}
\begin{document}

\maketitle

\begin{abstract}
We review an approach to observation-theory comparisons we call ``Taste-Testing."  In this approach, synthetic observations are made of numerical simulations, and then both real and synthetic observations are ``tasted"  (compared) using a variety of statistical tests.   We first lay out arguments for bringing theory to observational space rather than observations to theory space.  Next, we explain that generating synthetic observations is only a step along the way to the quantitative, statistical, taste tests that offer the most insight.   We offer a set of examples focused on polarimetry, scattering and emission by dust, and spectral-line mapping in star-forming regions.   We conclude with a discussion of the connection between statistical tests used to date and the physics we seek to understand.  In particular, we suggest that the ``lognormal" nature of molecular clouds can be created by the interaction of many random processes, as can the lognormal nature of the IMF, so that the fact that both the ``Clump Mass Function" (CMF) and IMF appear lognormal does not necessarily imply a direct relationship between them.

\end{abstract}

\firstsection % if your document starts with a section,
\section{Introduction}
Saying that one is working on ``comparing observations with simulations" sounds straightforward.  Alas, though, comparing astronomical experimental data (``observations") about star formation with relevant theory is much more difficult than [experiment]:[theory] comparisons in almost any other area of physics.  
The reasons are twofold. First, the theories involved are not typically ``clean" analytic ones, but instead require simulation to make predictions. And second, the measurements observers offer involve complicated-to-interpret ``fluxes" from various physical processes rather than more direct measures of physical parameters.

Let's say a theory makes a prediction about density.  In a laboratory, one fills a container of known interior volume and weight with the substance under study using some kind of simple mechanical device (e.g. a spoon), puts the container on a well-calibrated scale, measures a weight, subtracts the weight of the container, divides by the well-known gravitational constant and the known volume, and {\it voil\`a}, density.   In a star-forming region...not so simple.  One points a telescope of (sometimes poorly) known beam response at some patch of 2D sky and collects all the photons that come to a detector of (sometimes poorly known or variable) efficiency via the telescope during a well-measured period of time.  Such an observation gives the ``flux" (or lack thereof) of some quantity (e.g. thermal emission from dust, emission from a spectral line in gas, dust extinction).  Next, that flux needs to be converted either to a 2D ``column density" on the sky, or a 3D volume density (only possible using chemical excitation models for spectral lines).  In the case of a column density, the ``column" needs to be assigned a length to extract a volume density.  And, in either case, it is not clear what mixture of actual densities conspire along a line of sight to give the measured ``representative" volume density.

\begin{figure}[h]
% \vspace*{-2.0 cm}
%\begin{center}
 \includegraphics[width=5.5in]{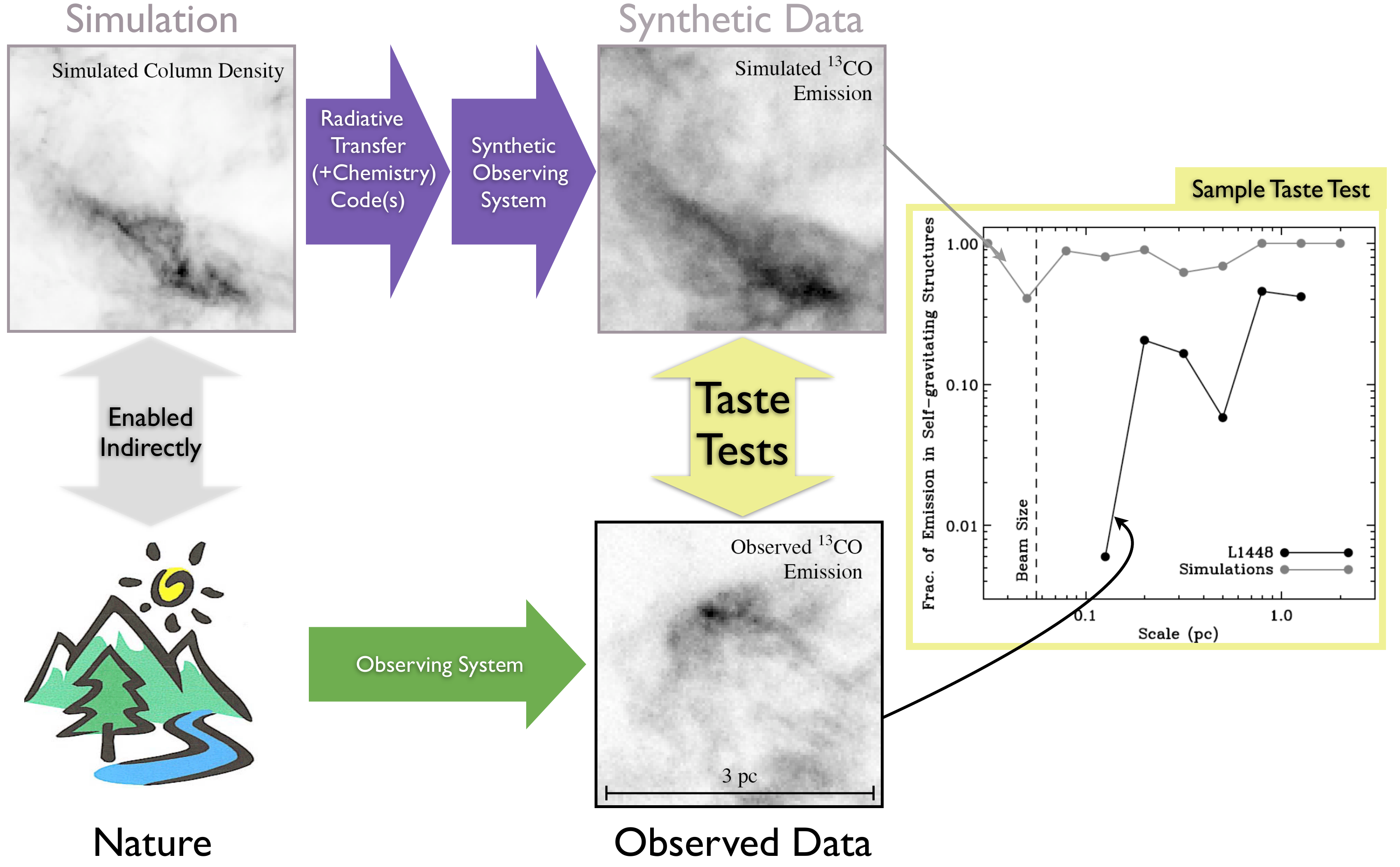} 
\vspace*{0.2 cm}
 \caption{The Taste-Testing Process.  The processes leading to measurements are represented by the green arrow for ``real" data and by the two purple arrows for synthetic data.  The box at right shows a comparison of the fraction of self-gravitating (according to a virial analysis) material as a function of scale in the simulations of \citet{Padoan:2006p711} and in the L1448 region of Perseus (see \citet {Rosolowsky:2008p53, GoodmanNat09} and \S 3, below.) }
   \label{fig1}
%\end{center}
\end{figure}

So, should we give up on observational measurements of physical parameters in star formation research?  Of course not.  But, we should consider that perhaps comparisons of observations and theory (primarily simulations) are best carried out in an ``observational space," where synthetic observations of theoretical output are made in order to enable more direct statistical comparison with ``real" (observational) data.   This comparison in observational space has been referred to in the past by some (including me) as ``Taste-Testing."  If one's goal is to reproduce the recipe for a great soup eaten in a restaurant, then a good course of action is to guess the ingredients based on dining experience (observation) and the processes based on cooking experience (physics) create the soup, take a look at it, and then if it seems to look and smell right, then finally ``taste" it.  Centrifuging the soup--as a theorist might try to do with observations in ``theorists'" space--to discern its ingredients might give you a basic chemical breakdown, but it would not tell you how to make it again.   Thus, we define a legitimate [observation]:[theory] taste test as one where statistical (taste) tests are applied in observer's (edible) space.  An overview of the process is shown in Fig.\,\ref{fig1}.

\section{Synthetic Observations}
In creating synthetic observations of star-forming regions, there are two conceptual steps (as shown by the purple arrows in Fig.\,\ref{fig1}).  First, the physical processes producing the spectral-spatial-temporal combination of photons to be observed must be modeled.  And, second, the response of the observing system to that flux of photons must be calculated.  In reality, each of these steps can be very difficult, and often one step is taken without the other.  For example, it is exceedingly difficult to model the molecular line emission from star-forming gas fully, as chemical, radiative, shock, magnetic, and thermal physics can all come into play.  So, researchers like \citet{Padoan:1999p1510}, \citet{Ayliffe07}, \citet{Offner:2008p1482}, \citet{Rundle2010} are to be forgiven when they create somewhat idealized ``synthetic" spectral line maps, not including the exact response of a particular telescope.   Other researchers all but skip the first step and carry out only the second, by running only highly idealized analytic models (rather than more complicated simulations) though telescope ``simulators," like the ALMA simulator (\url {http://www.cv.nrao.edu/naasc/alma_simulations.shtml}).   Some researchers, for example \citet{Krumholz:8p1543}, have already gone so far as to ``observe" numerical simulations  (e.g. of massive disks) with particular (synthetic) telescopes (e.g. ALMA, EVLA).  In principle, the ARTIST efforts (\url {http://www.astro.uni-bonn.de/ARC/artist/}) will be able to automate the process of creating synthetic observations for any ``input" theoretical model, once inputting complex numerical models becomes operationally straightforward.

\begin{figure}[h]
\vspace*{-0.5 cm}
 \includegraphics[width=5.2in]{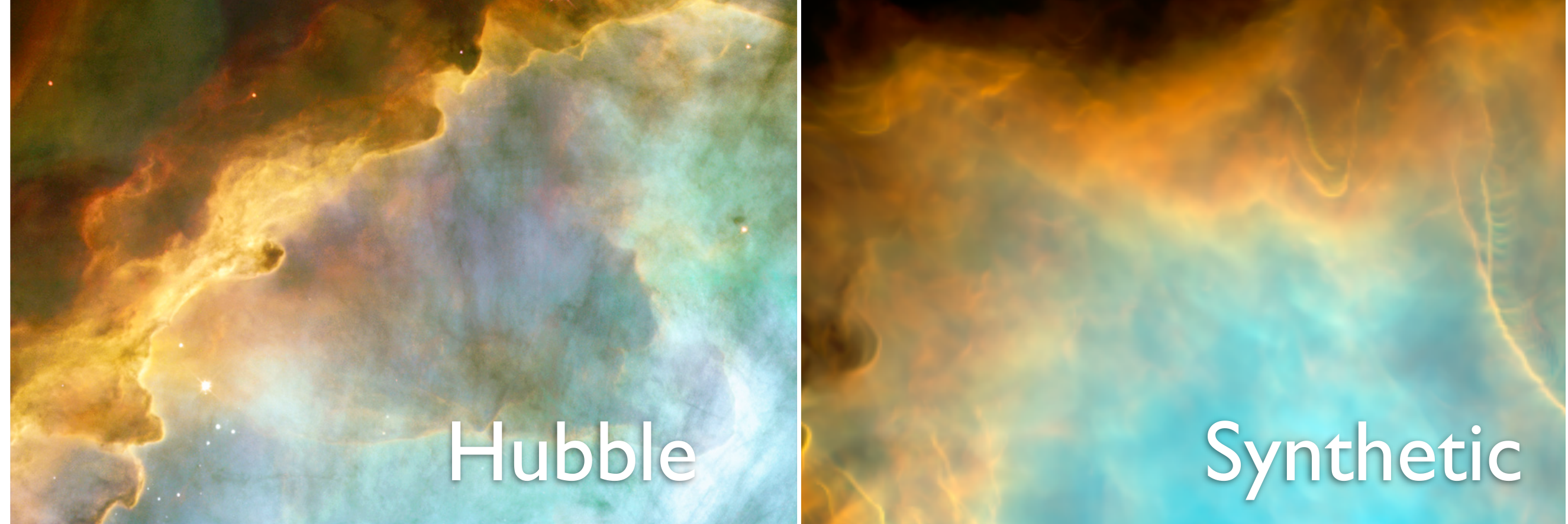} 
\begin{center}
 \caption{Will the real M17 please stand up?}
 \vspace*{-0.7 cm}
 \label{fig2}
 \end{center}
\end{figure}

Fig.\,\ref{fig2} shows what is certainly one of the most beautiful examples to date of a synthetic observation.  The left panel shows the [OIII, blue], H-$\alpha$ [green] and [SII, red] emission-line image (Hester, STScI-PRC03-13) of part of the M17 H II region from the Hubble Space Telescope, and the right panel shows a synthetic view through the same filters of a $512^3$ numerical simulation of \citet{Mellema:8p1553}.  

While perhaps less spectacularly beautiful, other synthetic products, such as spectral-line maps, polarization maps, dust emission, scattering, and extinction maps have been generated recently by many groups (see examples in Table 1), and more and more are being created as computational power increases.   All of these synthetic data products, created by what some groups call ``forward modeling," offer the opportunity for ``Taste Testing." It is critical to keep in mind that legitmate comparison requires the taste-testing step: the creation of synthetic observations is a key step toward comparison, but not a comparison itself.   Apple juice and iced tea can look very much like each other, but they do {\it not} taste the same.  The images in Fig.\,\ref{fig2} may {\it look} alike to our eyes, but are we being fooled by the matching color scheme and roughly correct morphology?  Or, is the similarity really great? 

\section{Statistical Comparisons (Taste-Tests)}

In order to quantify just {\it how similar} real and simulated observations are, we need powerful statistical measures, preferably of properties that are intuitively and or physically meaningful.  

Some comparative statistical tests are harder to pass than others, and the trick is to find a test that is ``hard enough" so that it cannot be fooled, while not being so hard as to require an (unrealistic) level of agreement.   For a trivial example of a test ``too easy" to pass, think about the intensity distribution (a histogram) of all pixels in a map. Intensity histograms might match for two maps, but so would a histogram for a totally unphysical map created by randomizing the position of every intensity pixel in either map.   Too easy.   

\begin{figure}[h]
% \vspace*{-2.0 cm}
%\begin{center}
 \includegraphics[width=5.5in]{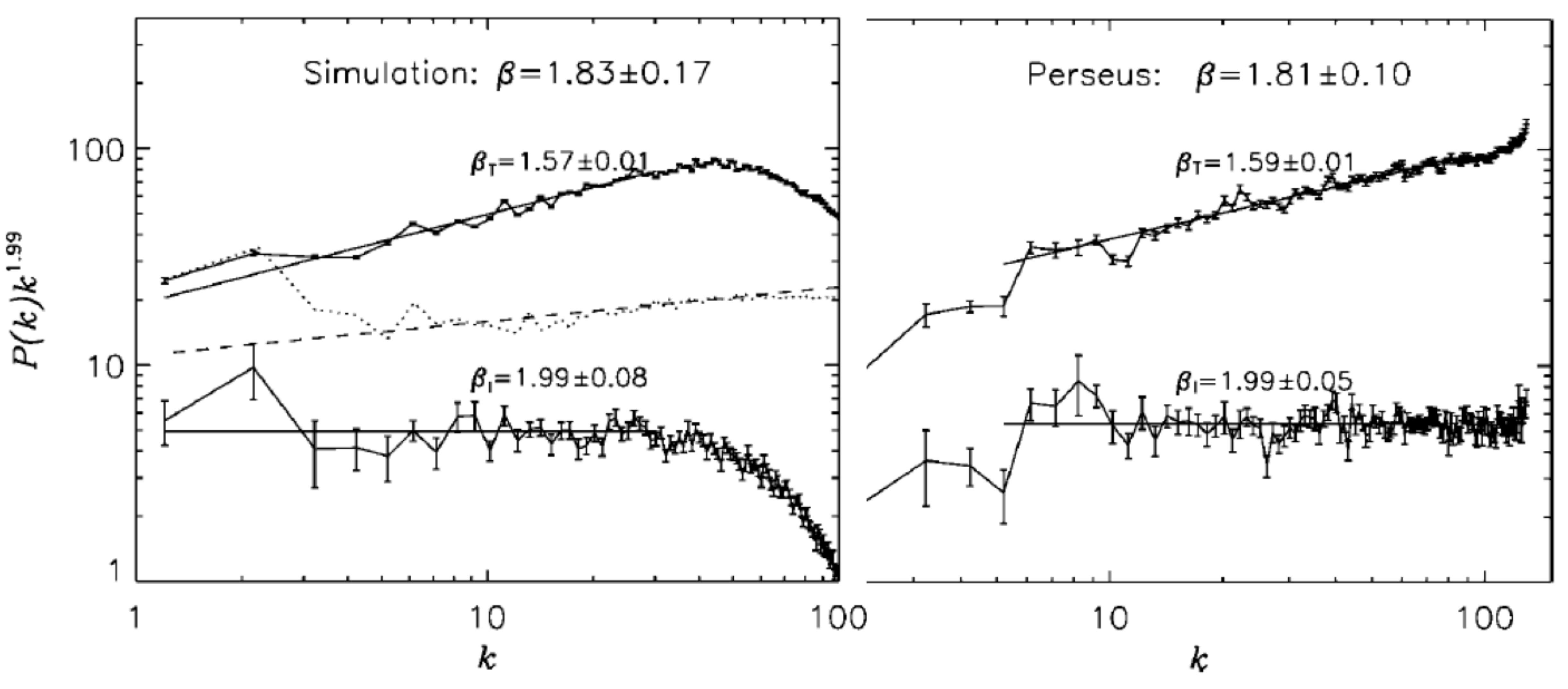} 
% \vspace*{-1.0 cm}
 \caption{Velocity Power Spectra of Synthetic (left) and Real (right) maps, based on Figures in \citet{Padoan:2006p711}}
   \label{fig3}
%\end{center}
\end{figure}

Many researchers make use of various forms of correlation functions, which take account of the distribution of some measured property at various real or transformed scales, as taste tests (see, for examples, \citet{Lazarian:3p1554, Padoan:5p1582, Brunt:2009p1558}, and references therein).  In our experience, passing any such test is often necessary, but not sufficient, evidence of plausible agreement. By way of example, consider Fig.\,\ref{fig3}, which shows velocity power spectra for synthetic and real $^{13}$CO maps of a star-forming region, from the work of \citet{Padoan:2006p711}  The agreement between the slopes of the power spectra shown is excellent, but further work, using harder-to-pass dendrogram-based tests, on the same data shows that the two data cubes used in the comparison are in fact quite different \citep{Rosolowsky:2008p53, GoodmanNat09}.

The right-most panel of Fig.\,\ref{fig1} shows a comparison of the L1448 sub-region of Perseus with a relevant sub-region of the same simulation used by \citet{Padoan:2006p711} to create the left panel of Fig.\,\ref{fig3}.  The hierarchy of both clouds is measured by constructing tree diagrams (dendrograms) that quantify the hierarchy of emission within the position-position-velocity space created by spectral line mapping. A virial parameter can be calculated for every point in the tree, using measured mass, size, and velocity dispersion.  The ``Sample Taste Test" panel of Fig.\,\ref{fig1} shows that the fraction of material with virial parameter below a ``self-gravitating" threshold does {\it not} depend on scale in the same way in both real and synthetic data cubes. (And, the disagreement is particularly off-putting because the majority of structures within the Padoan et al. simulation are apparently self-gravitating, even though self-gravity was not included as an input to the simulation.)  So, alas, while the synthetic observations of the simulations shown in \citet{Padoan:2006p711}  {\it pass} a velocity power spectrum comparison test, they {\it fail} the dendrogram-derived virial paramter test.  Note that this and any similar  failure to pass a taste test can be due to any step along the way through the top half of Fig.\,\ref{fig1}:  the discrepancy can stem from the simulation itself and/or from the radiative transfer and/or chemistry calculations and/or from assumptions about the telescope response.

\begin{table}
% \vspace*{-2.0 cm}
%\begin{center}
 \includegraphics[width=5.5in]{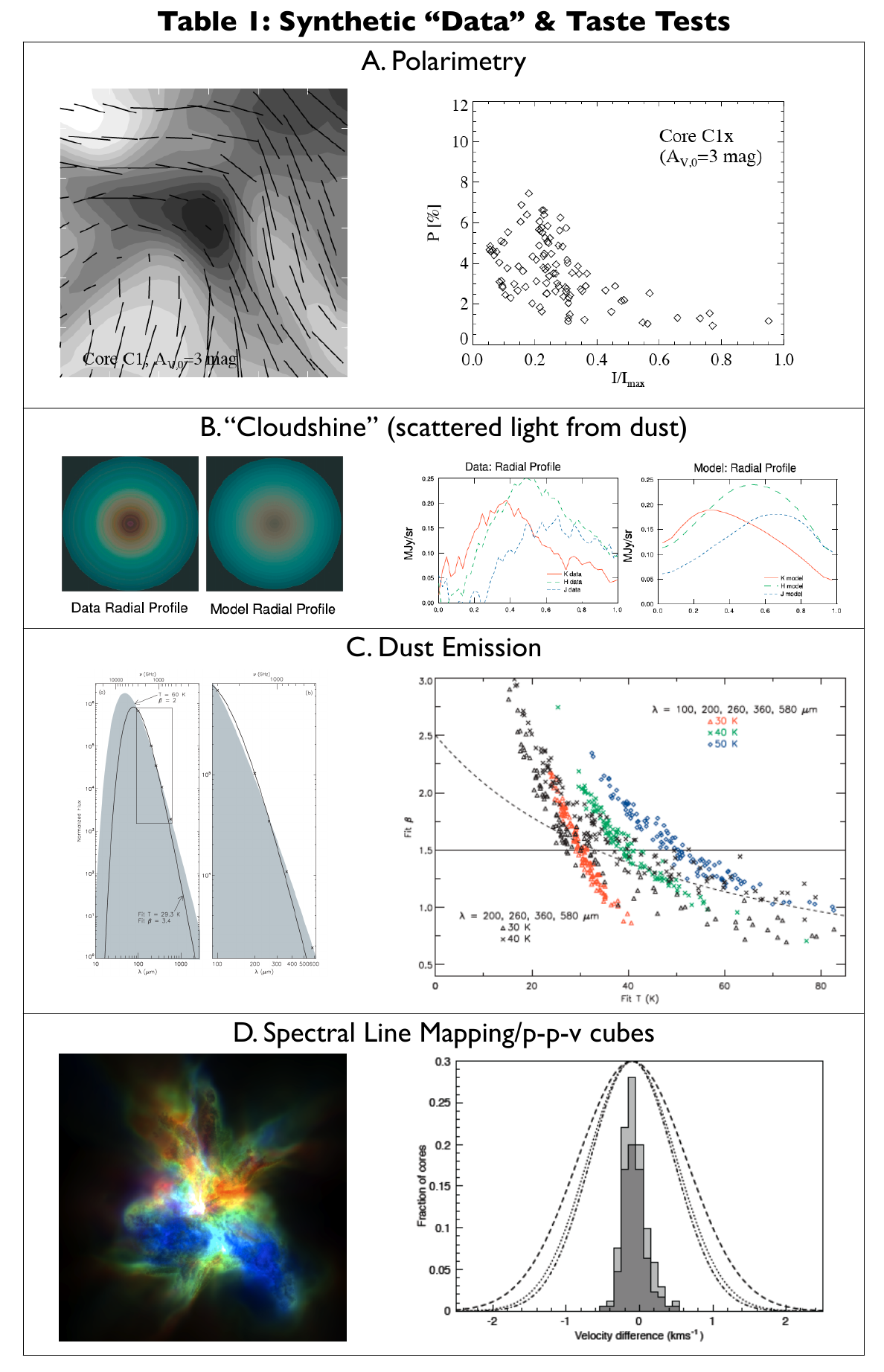} 
% \vspace*{-1.0 cm}
 \caption{Tasting Menu: Please refer to the text on the facing page for references and details. }
   \label{tab1}
%\end{center}
\end{table}

\section{Tasting Menu}
Since so much more ``Tasting" work has been done and is ongoing than what there is room to mention here, I shall take a ``Tasting Menu" approach to futher examples, much as a fine restaurant would. Table 1, as a Tasting Menu, offers four courses.  The text in the remainder of this section offers descriptions of each course.

For an {\it hors d'oeuvre}, consider {\bf ``A. Polarimetry".}  Synthetic polarization observations, giving insight into magnetic field structure, have been created by several authors (cf. \citet{2001ApJ...546..980O}).  The specific example shown in Table 1 shows the re-creation of a so-called ``polarization hole"  in a sub-mm dust emission map. In nearly all density regimes, polarization efficiency falls as the highest densities probed by a particular tracer are reached (see \citet{Goodman:8p1561, Matthews:8p1563, Lazarian:11p1560, Whittet:2p920} and references therein).  When SCUBA maps of dense cores were first made e.g. \citet{Matthews:11p1564}, the ``holes" this inefficiency causes in the center of maps seemed mysterious.  The work by \citet{Padoan:2001p1012} shown here demonstrates that a polarization map ({\it left panel, synthetic data}) has the same polarization-intensity behavior as the real data ({\it right panel, taste-test}).

For an appetizer, let's look at {\bf ``B. Cloudshine".}  Dark clouds are not ``dark" when imaged deeply enough.  The effect known as ``Cloudshine" was first observed in the NIR by \citet{Lehtinen:5p1565} then again by \citet{Foster:1p1566}, and recently by \citet{Steinacker:2p1567} in the MIR (a.k.a. ``coreshine").  \citet{Foster:1p1566} created very simple scattering models ({\it ``model radial profile" panels}) of the cloudshine from a dense core immersed in the interstellar radiation field and compared those with a dense core in L1451 ({\it ``data radial profile" panels}).  In the figure shown in Table 1, the taste-test is simply the agreement of the different colors' emission as a function of scale.  \citet{Padoan:1p1568} carried out more sophisticated modeling of the scattered light produced within a turbulent box and showed that column density can be recovered from NIR cloudshine at accuracies no worse than 40\% off (and typically much better) from reality.

For our main course, let's consider {\bf ``C. Emission"}. The properties of dust emission from star-forming regions will grow in prominence as a comparator in the near-term future in light of new Herschel results.  The particular example shown here comes from the work of \citet{Shetty:2009p55}.  In the {\it left panel} synthetic observations of the SED at Herschel-like wavelengths of a 60 K emissivity-modified ($\beta=2$) blackbody are shown: the solid line shows a fit to synthetic data where noise at an rms level of less than 10\% has been included at each wavelength, which gives quite far-off values of $T=29.3$ K and $\beta=3.4$.    The ``taste-test" comparison (showing $\beta$ as a function temperature) in the {\it right panel} shows of an ensemble (colored points)  of fit results akin to the single example shown at left, for slightly-noisy far-IR and sub-mm SEDs,  along with the observational findings (solid curve) of \citet{Dupac:2003p1570}.  Dupac et al. claim a physical origin for their measured``temperature-emissivity" correlation, while the Shetty et al. results show, using taste-tests, that the correlation could be an artifact of SED fitting degeneracies.

And, for dessert (the best part), let's look at {\bf ``D. Spectral Line Mapping/{\it p-p-v} cubes"}.  Spectral-line maps are both the richest sources of detailed information about molecular clouds, and the hardest to model correctly.  As explained above, since radiative transfer, chemistry, and telescope response all come into play, it is exceedingly difficult to confidently offer synthetic position-position-velocity ({\it p-p-v}) cubes for testing.  Thankfully, though, several research groups have been willing to tread in (or at least near) these treacherous waters of late \citep{Padoan:1999p1510, Ayliffe07, Offner:2008p1482, Kirk:7p1579, Tilley:8p1580, Rundle2010}.  In the example shown here, the {\it left panel} shows a map of HCO$^+$ (1-0) emission color-coded by velocity created by \citet{Rundle2010}.  The {\it right panel} shows a corresponding prediction of how narrow the distribution of core-to-core velocities (filled histograms) is with respect to the widths of various spectral lines in their environs (curves).  In particular, these results (cf. \cite{Ayliffe07}) show that even a simulation with significant amounts of competitive accretion can produce the kinds of small velocity offsets amongst features and tracers seen, for example, by \citet{Walsh:10p1583} and by \citet{Kirk:8p1810}.

\section{The Future: ``Tests of the Tests"}

A couple of years ago at a meeting in honor of Steve and Karen Strom, Frank Shu listened to my presentation of an early version of the L1448 results shown in Fig.\,\ref{fig1}.  He astutely commented:``that's nice, but {\it how about a Test of the Test}"?  What Frank meant by this question is critical for the future of observation-simulation comparisons, so allow me to explain. Sure, one can calculate a ``virial parameter," and use it as a statistical measure of two {\it p-p-v} cubes' similarity, but, is that virial parameter really a measure of the same physics (e.g. boundedness) that it would be in real ``{\it p-p-p v-v-v}" space?   Similarly, one can use CLUMPFIND to de-compose clouds statistically, but does it measure real structures within cloud, whose physical properties it is worth comparing? Does finding a lognormal column density distribution really uniquely imply anything about turbulence?  Let us consider these last three questions in turn.

\subsection{Virial Parameters and {\it p-p-v} Space }

Using the virial theorem--even in full, real, 3D ({\it p-p-p v-v-v}) space--is dangerous, in that key terms are often left out.  In recent years, virial analysis has been applied rather cavalierly in {\it p-p-v} space as well, thanks in part to Bertoldi and McKee (1992), who laid out the ground work for a ``virial parameter" that was to be a proxy for the significance of gravity as compared with turbulent pressure.  In recent work, \citet{Dib:5p1586} have shown that ignoring the subtleties of the virial theorm (e.g. surface energy) leads to incorrect conclusions about the boundedness of particular objects.  And, in work directly aimed at understanding the translation from features in {\it p-p-v} space to {\it p-p-p v-v-v} space \citep[c.f.][]{BallesterosParedes:5p1555, Smith:12p1587}, Shetty et al. have recently shown that the standard virial parameter \cite[e.g.][]{Bertoldi:8p1584} is really only ``safe" for very simple, isolated, objects, and that the more crowded a region becomes, the worse an approximation the virial paramter is. Thus, new work seeking perhaps more robust measures--in {\it p-p-v} space of the role of gravity and fundamental forces/processes in molecular clouds--is sorely needed.

\subsection{Structures, Real or Statistical?}

We heard much at the meeting about ``clump mass spectra" and about the ``filamentariness" of molecular clouds.  However, unless one allows for the possibility of ``clumps" {\it within} ``filaments," these measures are topologically inconsistent.  Thus, measures of cloud structure that do {\it not} allow for hierarchy are likely to produce unphysical structures.   For example, when CLUMPFIND \citep{Williams:6p1588} is applied in very crowded regions, where hierarchy and overlap are important, not only will spurious, unphysical, structures be cataloged, the exact structures found will depend sentitively on CLUMPFIND's input parameters \citep{Pineda:2009p46}.  So, while CLUMPFIND can be validly used as a purely statistical comparative test, the na\"ive {\it interpretation} of its output as giving a mass spectrum of structures that can directly form individual or small groups of stars, is not physically sensible.  (Note that in cases where extended structure is removed before CLUMPFIND is applied  \citep[e.g.][]{Alves:1p1589},  the resulting Clump Mass Function (CMF) is then more valid, but it is still reliant on the exact prescription for removing ``extended" structure.)

What is needed instead of CLUMPFIND?  More unbiased algorithms like the dendrogram (tree) approach blatantly advertised above are a good first step.  But, even these more sophisticated segmentations then need measures of the connections between physically-relevant parameters and purely statistical ones.   For example,  in {\it p-p-v} space, can we find use new segmentation algorithms like dendrograms to help find a relationship between how ``filamentary" a region is, and how ``bound" those filaments, or perhaps the cores within them, are?

\subsection{Lognormals are Not Enough}

Many authors have recently shown that when a ``large enough" volume of a molecular cloud is sampled, a lognormal column density distribution will be found (\cite{Goodman:2009p48, Wong:5p1593, Froebrich:8p1591, Kainulainen:12p1596, Lombardi:3p1597}).  The existence of such lognormals is certainly {\it consistent} with the predictions of many turbulence simulations \citep[e.g.][]{VazquezSemadeni:3p1595, 2001ApJ...546..980O, Wada:2001p1594}.  However, a Bonnor-Ebert sphere sampled appropriately will also give a lognormal density distribution \citep{2010MNRAS.tmp.1217T}!  And, in fact, the central-limit theorem shows that lognormal behavior will result from a confluence of nearly any set of interacting random processes.   Thus, measuring lognormal behavior is again ``necessary but not sufficient" on its own as a descriptive statistic.   It may be, though, that the particular properties of a lognormal (e.g. its width, its mean, and/or over what range it holds) may be more discriminating.

Just because the IMF and CMF are both consistent with lognormals does {\it not} mean one comes definitively directly from the other. As \citet{Swift:2008p1765} show, assuming very different core-to-star fragmentation/multiplicity schemes can,  at the level of detail we can measure them today, produce the ``same" IMF from the ``same" CMF.  It is indeed possible that the CMF could easily be generated by the random (even turbulent!) processes within molecular clouds on large scales, while the IMF is {\it simultaneously} generated by the randomness \citep{Adams:6p1598} that takes place on smaller scales within clusters and fragmenting cores.   
\subsection{Summary}
It is clear that some ``taste" tests are harder to fool than others, so clearly, we will seek and prefer the tough ones as discriminators. Our community's  next step will be to begin to seriously explore which tests are best for discerning which physical characteristics, as this is not yet very clear.  We should think of this process as designing artificial ``instruments" capable of measuring the interstellar equivalents of culinary measures like acidity, texture, and salinity: only, our instruments will measure properties like ionization,  magnetic beta, vorticity, etc.  As the computational power available for simulation and the size of observational databases increase, we will all need to learn how to become tough but fair food critics, and good chefs.

%\begin{thebibliography}{}
%\bibitem[Anders \& Zinner (1993)]{AndersZinner93}
%{Anders, E., \& Zinner, E.} 1993, 
%\textit{Meteoritics}, 28, 490

%\end{thebibliography}

\bibliography{goodman}{}
\bibliographystyle{apj}
\end{document}